\documentclass[aps,amsmath,amssymb,physrev,reprint]{revtex4-2}

\usepackage{graphicx}
\usepackage{dcolumn}
\usepackage{bm}
\graphicspath{{./figures/}}

\usepackage[total={6.5in,9.75in}, top=0.9 in, left=0.9in, includefoot,
]{geometry}
\usepackage{color}

\usepackage[normalem]{ulem}

\usepackage{tikz,xcolor,hyperref}

\definecolor{lime}{HTML}{A6CE39}
\DeclareRobustCommand{\orcidicon}{
	\begin{tikzpicture}
	\draw[lime, fill=lime] (0,0) 
	circle [radius=0.16] 
	node[white] {{\fontfamily{qag}\selectfont \tiny ID}};
	\draw[white, fill=white] (-0.0625,0.095) 
	circle [radius=0.07];
	\end{tikzpicture}
	\hspace{-2mm}
}
	
\foreach \x in {A, ..., Z}{\expandafter\xdef\csname orcid\x\endcsname{\noexpand\href{https://orcid.org/\csname orcidauthor\x\endcsname}
			{\noexpand\orcidicon}}
}

\begin{document}

\preprint{APS/123-QED}

\title{Elastic turbulence in straight confined geometries}

\author{Giulio Foggi Rota \orcidA{}$^1$}
\thanks{These authors contributed equally.}
\author{Daulet Izbassarov \orcidB{}$^{2,3}$}
\thanks{These authors contributed equally.}
\author{Aditya Tekriwal$^1$}
\thanks{Currently at: Department of Physics, University of Oxford, Oxford OX13PU, United Kingdom.}
\author{Marco Edoardo Rosti \orcidC{}$^1$}
\email[E-mail for correspondence: ]{marco.rosti@oist.jp}
\affiliation{
$^1$ Complex Fluids and Flows Unit, Okinawa Institute of Science and Technology Graduate University, 1919-1 Tancha, Onna-son, Okinawa 904-0495, Japan.\\ $^2$ Theoretical Sciences Visiting Program, Okinawa Institute of Science and Technology Graduate University, 1919-1 Tancha, Onna-son, Okinawa 904-0495, Japan.\\ $^3$ Finnish Meteorological Institute, Erik Palmenin aukio 1, Helsinki 00560, Finland.\\ 
}

\begin{abstract}
\noindent
Elastic turbulence (ET) in straight geometries has been extensively studied in idealised configurations such as periodic shear and planar channel flow, yet experiments are typically restricted to confined systems like ducts and pipes. While chaotic elastic fluctuations are observed in ducts, their onset in pipes is debated due to {geometric} constraints. Here we report sustained chaotic dynamics in both geometries and show they correspond to the same ET state identified in idealised setups. Our results establish ET as a geometry-independent attractor and point to a distinct transition mechanism in pipes.
\end{abstract}

\maketitle


Fluid elasticity can sustain chaotic flow regardless of inertial effects, in the state termed \textit{Elastic Turbulence} (ET) \cite{groisman-steinberg-2000}.
In laminar flows of viscoelastic fluids along curved streamlines, perturbations grow through the coupling between normal and radial stresses \cite{bird-1987,pakdel-mckinley-1996,groisman-steinberg-2001-1}, eventually leading to fully developed and sustained ET.
ET is nevertheless attained also in flow configurations where the streamlines of the laminar state are straight lines \cite{pan-etal-2013}, e.g.\ planar shear (Couette) \cite{beneitez-page-kerswell-2023} and planar channels \cite{lellep-linkmann-morozov-2024,foggirota-etal-2024-3}.
In such geometries, where curvature of the base flow is absent, the transition mechanism is fundamentally different.


{
In planar channels, although the full pathway from laminar flow to three-dimensional ET remains unresolved, the onset of a centre-mode instability \cite{page-dubief-kerswell-2020,khalid-etal-2021,buza-etal-2022}, followed by its modulation in the spanwise direction \cite{lellep-linkmann-morozov-2023}, is believed to initiate transition.
This secondary instability exhibits two defining properties: reflectional symmetry about the centre-plane, and non-vanishing streamwise velocity exactly on that plane.
}


{
In circular pipes, where the centre-mode instability was first isolated \cite{garg-etal-2018-1}, the symmetry element reduces to the one-dimensional centreline.
The two halves of any diameter through this centreline correspond to azimuthal angles differing by $\pi$; translated to the pipe, the reflectional symmetry characteristic of the channel mechanism would therefore require a modulation on even azimuthal wavenumbers \cite{kerswell-davey-1996} -- to our knowledge, never isolated thus far.
The second defining property of the channel mechanism is even more restrictive: in the pipe, any perturbation varying with the azimuthal angle must vanish at the centreline to avoid becoming multi-valued, leaving axisymmetric perturbations (i.e., with null azimuthal wavenumber) as the only ones that can be finite there.
Since the pipe centre-mode itself is indeed purely axisymmetric \cite{garg-etal-2018-1,chaundhary-etal-2021}, it thus becomes apparent that, if pipe flow of viscoelastic fluid transitions to ET, it must proceed through a route different from that of channel flow.
}

The constraints discussed above, together with the lack of clear experimental evidence of fully developed ET in pipes (only elasto-inertial turbulence has been reported \cite{choueiri-etal-2021, kumar-graham-2026}), raises a central question regarding the very existence and nature of ET in such geometry.
Confined flow configurations such as pipes and ducts are of particular importance, as they are experimentally accessible and directly relevant to medical and engineering applications.
While fully chaotic viscoelastic flows are commonly observed in duct experiments at low inertia \cite{pan-etal-2013,jha-steinberg-2021,li-steinberg-2023}, pipe experiments only report weakly chaotic and three-dimensional states near the centre-mode instability threshold \cite{choueiri-etal-2021}, without conclusively distinguishing between incipient subcritical transition and nonlinear saturation of the supercritical mode.

\begin{figure*}
   \includegraphics[width=\textwidth]{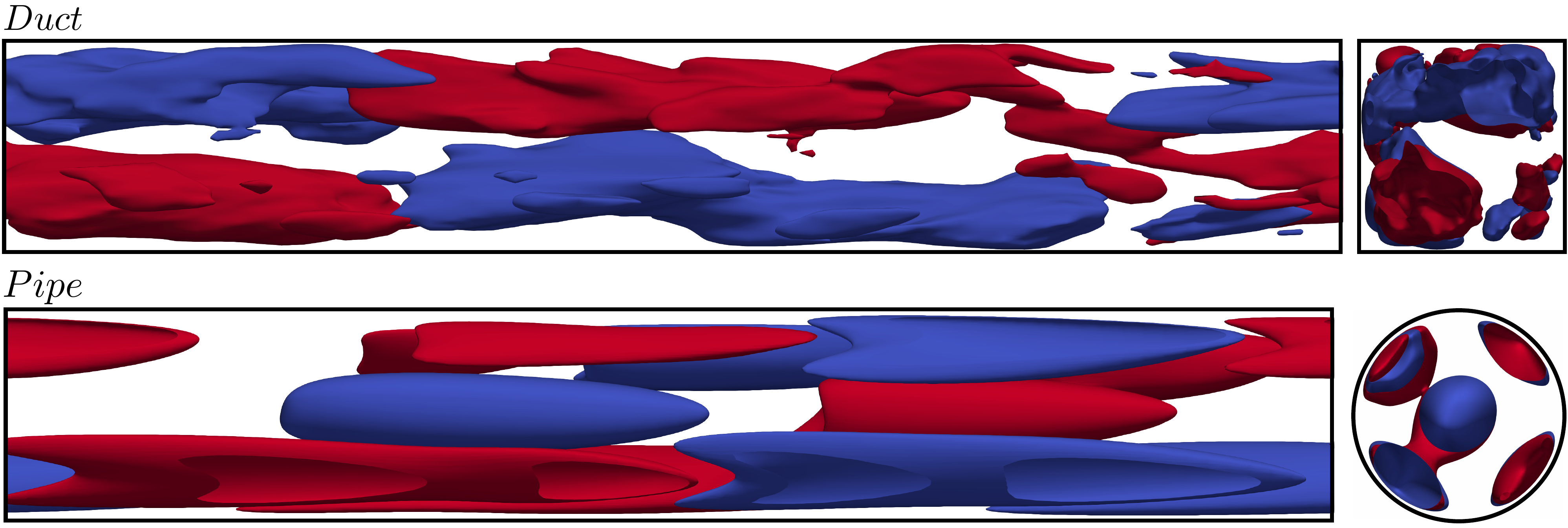}
   \caption{\textbf{Side and back views} (left/right) of the streamwise velocity fluctuations iso-surfaces in the duct (top) and pipe (bottom) geometries. Levels are set at $u^\prime=\pm 8 \cdot 10^{-4} U$ in the duct, and $u^\prime=\pm 1.6 \cdot 10^{-2} U$ in the pipe. The flow goes from left to right in the side views, and enters the page in the back views.}
   \label{fig:vis}
\end{figure*}

In this Letter we i) report the attainment of fully chaotic and sustained elasticity-driven motions of viscoelastic fluids in confined geometries (a duct and a pipe, in Fig.~\ref{fig:vis}), and ii) demonstrate that these states coincide with the ET regime \cite{zhang-etal-2025, foggirota-etal-2026} previously isolated in tri-periodic configurations \cite{singh-etal-2024}, planar channels \cite{lellep-linkmann-morozov-2024,foggirota-etal-2024-3}, and jets \cite{soligo-rosti-2023}.
Our results imply that, although geometry constrains the admissible transition pathways, it does not constrain the final ET state.
Pipe flow must therefore access ET through mechanisms distinct from those operating in channels, while the fully developed ET state itself emerges as a geometry-independent chaotic attractor.
 

The velocity field $\boldsymbol{u}$ is governed by the Navier--Stokes equations together with the incompressibility constraint. 
The dynamics of the fluid microstructure $\boldsymbol{C}$ are described by the Oldroyd-B model \cite{oldroyd-1950}. 
The flow thus evolves according to
\begin{equation}
\rho \left( \partial_t \boldsymbol{u} +  \left( \boldsymbol{u} \cdot \boldsymbol{\nabla} \right) \boldsymbol{u} \right)=-\boldsymbol{\nabla} p+ \mu_f \nabla^2 {\boldsymbol{u}} +\frac{\mu_p}{\tau} \boldsymbol{\nabla} \cdot \boldsymbol{C},
\label{eq:1}
\end{equation}
\begin{equation}
\boldsymbol{\nabla} \cdot \boldsymbol{u} =0,
\label{eq:2}
\end{equation}
\begin{equation}
\partial_t \boldsymbol{C} + \boldsymbol{u} \cdot \boldsymbol{\nabla} \boldsymbol{C} = \boldsymbol{C} \cdot \boldsymbol{\nabla}  \boldsymbol{u} + \boldsymbol{\nabla} \boldsymbol{u}^{T} \cdot \boldsymbol{C} - \frac{\boldsymbol{C}-\textbf{I}}{\tau},
\label{eq:3}
\end{equation}
where $\rho$ is the fluid density, $\mu_f$ the solvent viscosity, $\mu_p$ the viscosity contribution associated with the microstructure, and $\tau$ its relaxation time.
These equations are solved numerically using our extensively validated solver \textit{Fujin} \cite{rosti-2026} (\url{https://www.oist.jp/research/research-units/cffu/fujin}), in square duct and circular pipe geometries. 
Further details on the numerical methods, grid resolutions, domain sizes, and flow initialisation are provided in the End Matter.

The flow is characterised by the Reynolds number ($Re$), quantifying the relative importance of inertial and viscous effects, and the Deborah number ($De$), comparing the micro-structural relaxation time with the flow time scale. 
These are defined as
\begin{equation}
Re=\frac{\rho U \mathcal{L}}{\mu_f+\mu_p}
\qquad \textrm{and} \qquad
De=\frac{\tau U}{\mathcal{L}},
\end{equation}
where the velocity scale $U$ is the mean flow speed (maintained constant throughout simulations) and the length scale $\mathcal{L}$ is a macroscopic characteristic of the cross-section (the half-edge length for the duct, and the radius for the pipe). 
All simulations are performed at $Re=50$ and $De=50$, with a fixed viscosity ratio $\beta=\mu_f/(\mu_f+\mu_p)=0.9$.
{Although the chosen value of $Re$ is finite and larger than unity, the resulting elasticity number $E=De/Re=1$ has been demonstrated to characterise pure ET in planar channels~\cite{foggirota-etal-2024-3} and Taylor-Couette flows~\cite{zhang-etal-2025}.}


At steady state, both duct and pipe flows exhibit sustained velocity fluctuations characterised by the coexistence of large-scale motions and localised dynamics. 
The two configurations, however, differ markedly. 
Similarly to what is reported in channels \cite{foggirota-etal-2024-3}, coherent structures are observed at the duct centre and finer-scale motions at the walls, reminiscent of the centre-mode \cite{page-dubief-kerswell-2020,buza-etal-2022} and of the unsaturated polymer diffusive instability (PDI) \cite{beneitez-page-kerswell-2023,beneitez-etal-2024-2, suryaphanitej-kumarmohanty-shankar-2025, pandey-shankar-2025} (visualisations in the End Matter). 
Superimposed on these, we nevertheless identify streak-like structures at intermediate distances from the centre, meandering between the corners and the face centres (Fig.~\ref{fig:vis}, top), never reported in prior studies to the best of our knowledge and specific to the duct geometry. 
In the pipe, centre-localised structures akin to the centre-mode \cite{garg-etal-2018-1,chaundhary-etal-2021} are observed along with large-scale motions localised at the wall, evocative of the saturated PDI (Fig.~\ref{fig:vis}, bottom, and visualisations in the End Matter).
Remarkably, velocity fluctuations in the pipe do not comply with the axial symmetry of the geometry (also visible for the Q-criterion in the End Matter), hinting towards the attainment of chaotic dynamics rather than a pure and saturated centre-mode state.

\begin{figure}
   \includegraphics[width=\linewidth]{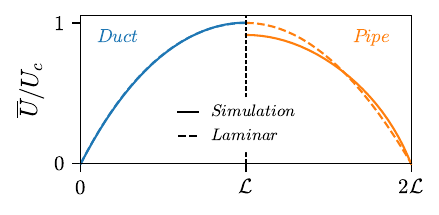}
   \caption{\textbf{Mean profiles of the streamwise velocity} for the duct (left) and the pipe (right), normalised with the centreline velocity of the laminar solution, $U_c$. Laminar profiles are shown as dashed lines for comparison. {Profiles are extracted through the cross-section centre, orthogonal to the faces in the duct.}}
   \label{fig:profiles}
\end{figure}

\begin{figure}
   \includegraphics[width=\linewidth]{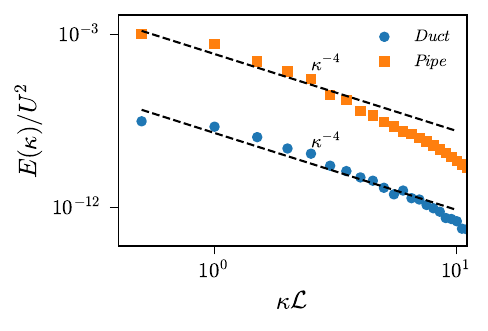}
   \includegraphics[width=\linewidth]{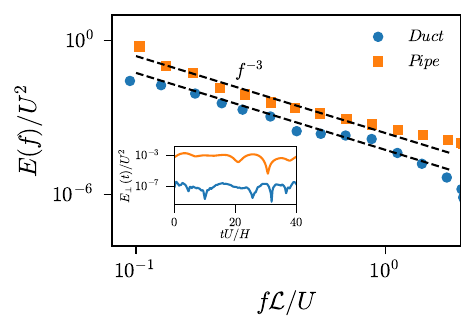}
   \caption{\textbf{Wavenumber and frequency spectra} (top/bottom) of the fluctuating energy $E$ at the centreline of their respective geometry. Wavenumber spectra are computed along the streamwise direction. In the bottom panel inset we show extracts from the time-histories of the cross-stream energy $E_\perp$, measured with a centreline probe.}
   \label{fig:spectra}
\end{figure}

Differences between the two setups extend also to the mean flow (Fig.~\ref{fig:profiles}). 
In the duct, the face-centred streamwise velocity profile $\overline{U}(y)$ remains close to the laminar parabola.
In the pipe, streamwise-averages of the streamwise velocity at specific time instants do not comply with axial symmetry and attain their maxima away from the centre. 
The time- and azimuthally-averaged velocity profile, $\overline{U}(r)$, thus exhibits a plug-like shape, with depleted shear ($\sim \partial \overline{U}/\partial r$) in the central region relative to laminar flow. 
This reflects a strong modulation of the mean flow by coherent structures, which are significantly more energetic in the pipe, where fluctuation levels exceed those in the duct by orders of magnitude (Fig.~\ref{fig:spectra}).
Such divergence might appear consistent with the curvature effect introduced previously, becoming relevant once {streamlines are curved} by the developing centre-mode and not hindered by flat boundaries.
Despite the macroscopic differences above, we now demonstrate that the underlying turbulent state is identical in the two systems and corresponds to ET.

 \begin{figure}
   \includegraphics[width=\linewidth]{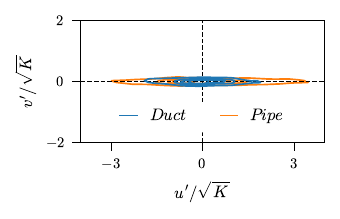}
   \caption{\textbf{Joint probability density functions} of the normalised streamwise ($u^\prime$) and wall-normal ($v^\prime$) velocity fluctuations at $\mathcal{L}/2$ from the wall in the duct (orange) and the pipe (blue). Normalisation relies on the mean turbulent kinetic energy $K$ in the considered geometry, and isolines are drawn at $\{0.1, 0.3, 0.5, 0.7\}$ the function maxima.}
   \label{fig:jpdf}
\end{figure}

Streamwise and wall-normal velocity fluctuations are fully de-correlated (Fig.~\ref{fig:jpdf}) and thus give an effectively null contribution to the Reynolds stresses: a defining feature of ET \cite{foggirota-etal-2024-3}.
Chaotic motions can therefore develop decoupled from the mean flow, allowing for both the near-laminar profile of the duct and the plug-like profile of the the pipe, which we impute to the more intense viscoelastic stresses developed in that case.

Velocity fluctuations permeate all flow scales and carry the spectral signature of ET: the steep $-4$ power-law exponent for the spatial decay of the fluctuating energy spectrum $E$ (Fig.~\ref{fig:spectra}, top).
As demonstrated in former work \cite{foggirota-etal-2026}, spatial and temporal scalings (Fig.~\ref{fig:spectra}, bottom) differ due to the breakdown of Taylor's hypothesis.
The sustained and developed nature of the ET state attained is confirmed by the time-histories of the kinetic energy $E_\perp$ associated with cross-stream velocity fluctuations at the centreline (inset of Fig.~\ref{fig:spectra}, bottom), oscillating about a finite value. In the pipe, these fluctuations necessarily break axial symmetry and therefore cannot be attributed to the centre-mode alone, demonstrating the presence of fully developed chaotic dynamics in addition to it.
The time-history reported for the duct covers the whole range of accumulated statistics after initial transients elapsed. Convergence is assessed by continuing the simulation on a coarser grid for further $400 \mathcal{L}/U$, leading to a slight depletion of $E_\perp$ due to enhanced diffusivity but otherwise no appreciable variation of the flow state attained.
For the pipe, instead, changing the grid (hence the stress diffusivity) appears to alter flow structures and leads to a more pronounced plug-like velocity profile, while turbulent statistics remain unchanged. Reported measurements are thus accumulated on a fine grid over $100 \mathcal{L}/U$, at the limit of what is numerically feasible with our tools.
Full details on the above variations, along with a further pipe simulation extending for $2400 \mathcal{L}/U$ on a coarser grid, are discussed in the End Matter.

 \begin{figure}
   \includegraphics[width=\linewidth]{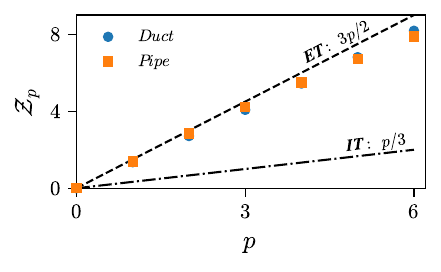}
   \caption{\textbf{Scaling exponents} of the $p^{th}$-order structure functions $\Sigma_p$ at sub-viscous scales, at the duct and pipe centrelines. Theoretical predictions without intermittency are reported as dashed/dash-dotted lines for ET/IT, respectively.}
   \label{fig:sexp}
\end{figure}

The ultimate confirmation that the chaotic dynamics described here are those of ET is provided by the scaling of the structure functions of the symmetric three-point velocity increment, $\Sigma_p=\langle \delta^2 u(r)^p \rangle$, where $\delta^2 u(r) = \left[\boldsymbol{u}(\boldsymbol{x}+\boldsymbol{r}) -2 \boldsymbol{u}(\boldsymbol{x}) + \boldsymbol{u}(\boldsymbol{x}-\boldsymbol{r})\right] \cdot \hat{\boldsymbol{r}}$ {and angled brackets denote ensemble-averaging}. 
This definition removes leading-order analytic contributions, so that at sub-viscous scales $\Sigma_p(r) \sim r^{\mathcal{Z}_p}$, with $\mathcal{Z}_p \sim 3p/2$ predicted for ET \cite{singh-etal-2024}, in contrast to $\mathcal{Z}_p \sim p/3$ in inertial turbulence (IT).
Our measurements (Fig.~\ref{fig:sexp}) follow the ET scaling in both duct and pipe. 
At high orders, the exponent saturation below $3p/2$ is consistent with the appearance of intermittency \cite{singh-etal-2024, garg-rosti-2025}.
{Finally, the good agreement of our measurements with the established fingerprints of ET verified above confirms the sub-dominant role of inertial effects relative to the elastic ones in the small, finite-$Re$ regime considered.}


The results presented here, {limited to the rheological model considered,} demonstrate that elastic turbulence can be sustained in confined geometries even in the absence of curvature in the base flow, and that the resulting chaotic state coincides with the same ET regime previously identified in tri-periodic \cite{berti-boffetta-2010,dzanic-from-sauret-2022,singh-etal-2024,lewy-kerswell-2025-1}, planar \cite{quin-arratia-2017,varshney-steinberg-2019, khalid-shankar-subramanian-2021, beneitez-etal-2024-2,lellep-linkmann-morozov-2024,foggirota-etal-2024-3}, and free-shear configurations \cite{yamani-etal-2021,yamani-etal-2023,soligo-rosti-2023}. 
ET therefore constitutes a unique dynamical attractor independent of flow geometry. 
{
This has direct implications for viscoelastic pipe flow. 
As discussed above, a pipe cannot reproduce the transverse modulation identified in channel flow \cite{lellep-linkmann-morozov-2023} if a non-zero streamwise velocity at the symmetry element is one of its defining properties, since the pipe's centreline -- unlike the channel's centre-plane -- cannot support such a disturbance away from full axial symmetry.
The attainment of ET in pipes thus implies the existence of an alternative route to chaos, possibly mediated by a secondary instability distinct from the one identified in channels, or by a mechanism akin to a bypass transition \cite{itoh-1977, kerswell-2005, avila-barkley-hof-2022}. 
Clarifying the nature of this transition remains an open problem.
}

Despite convergence towards the same chaotic attractor, the flow retains clear signatures of the underlying geometry. 
In the duct, we observe novel streak-like structures at intermediate locations and unsaturated PDI manifestations at the walls \cite{beneitez-page-kerswell-2023,beneitez-etal-2024-2}. 
In the pipe, by contrast, the PDI manifests in its saturated form.
Coherent structures are overall more energetic in the pipe, consistent with enhanced mean-flow modulation and the emergence of a time-averaged, plug-like velocity profile. 
These findings raise questions for future work. 
First, what mechanism underlies the transition to ET in pipe flow in the absence of the ``channel route"? 
Second, what is the role of coherent structures within ET {(e.g., the emerging centre-mode and PDI manifestations)}, and how do they couple to the background chaotic fluctuations and to the mean flow? 
Addressing these questions will be essential to develop a unified understanding of elasticity-driven turbulence in complex flows.

\begin{acknowledgments}
The research was supported by the Okinawa Institute of Science and Technology Graduate University (OIST) with subsidy funding to M.E.R. from the Cabinet Office, Government of Japan. The authors acknowledge the computer time provided by the Scientific Computing and Data Analysis section of the Core Facilities at OIST, and the computational resources offered by the HPCI System Research Project with grant hp260009 . 
\end{acknowledgments}

\section*{Data Availability Statement}
All information needed to evaluate the conclusions of this Letter is present in the main text and/or the End Matter. 
Data required to reproduce the figures are available on the website of the Complex Fluids and Flows Unit at OIST(\url{https://www.oist.jp/research/research-units/cffu/publications/publication-data}). Further material is available from the authors upon reasonable request.

%

\section*{End Matter}

\subsection*{Computational details}

Our in-house solver \textit{Fujin} \cite{rosti-2026} (\url{https://www.oist.jp/research/research-units/cffu/fujin}) uses a staggered, uniform Cartesian grid and discretises the governing equations (Eq.~\ref{eq:1}-\ref{eq:3}) in space with a second-order central finite-difference scheme. Time advancement is performed using a second-order Adams–Bashforth method, combined with a fractional-step approach \cite{kim-moin-1985}.
At each time step, the pressure field enforcing incompressibility is obtained via an efficient spectral Poisson solver, while domain decomposition and parallelisation are handled through the \textit{2decomp} library together with the message passing interface (MPI).
The evolution equation for the conformation tensor $\mathbf{C}$ requires particular attention: we adopt a logarithmic formulation \cite{fattal-kupferman-2005,devita-etal-2018} to mitigate the well-known high-Deborah-number instability, and employ a high-order weighted essentially non-oscillatory (WENO) scheme \cite{sugiyama-etal-2011} for the discretisation of the upper-convected derivative on the left-hand side.
This strategy avoids the need for an explicit stress-diffusion term, and thus no additional boundary conditions are required for $\mathbf{C}$ \cite{beneitez-etal-2024}.
For the velocity field, no-slip and no-penetration conditions are imposed at the walls, while periodic boundary conditions are applied in the homogeneous streamwise direction.
The curved pipe wall is obtained with an Immersed Boundary Method (IBM) \cite{hori-rosti-takagi-2022}.

\begin{figure}
   \includegraphics[width=\linewidth]{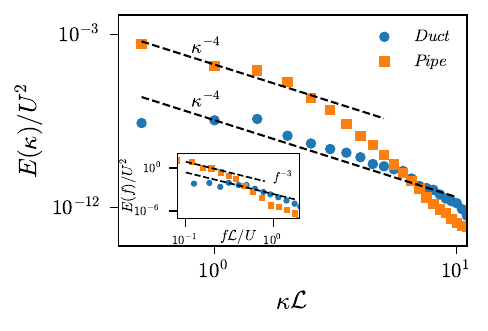}   
   \caption{\textbf{Wavenumber/frequency spectra of the fluctuating energy from our duct and pipe simulations on a coarser grid} (main axes/inset, respectively), at the centreline of the geometry. Wavenumber spectra are computed along the streamwise direction. The same scaling behaviours reported on the finer grid (Fig.~\ref{fig:spectra}) are achieved, even thought the scaling range in the pipe appears shrunk, likely due to the loss of finer-scale motions.}
   \label{fig:spectraUR}
\end{figure}

\begin{figure*}
   \includegraphics[width=\textwidth]{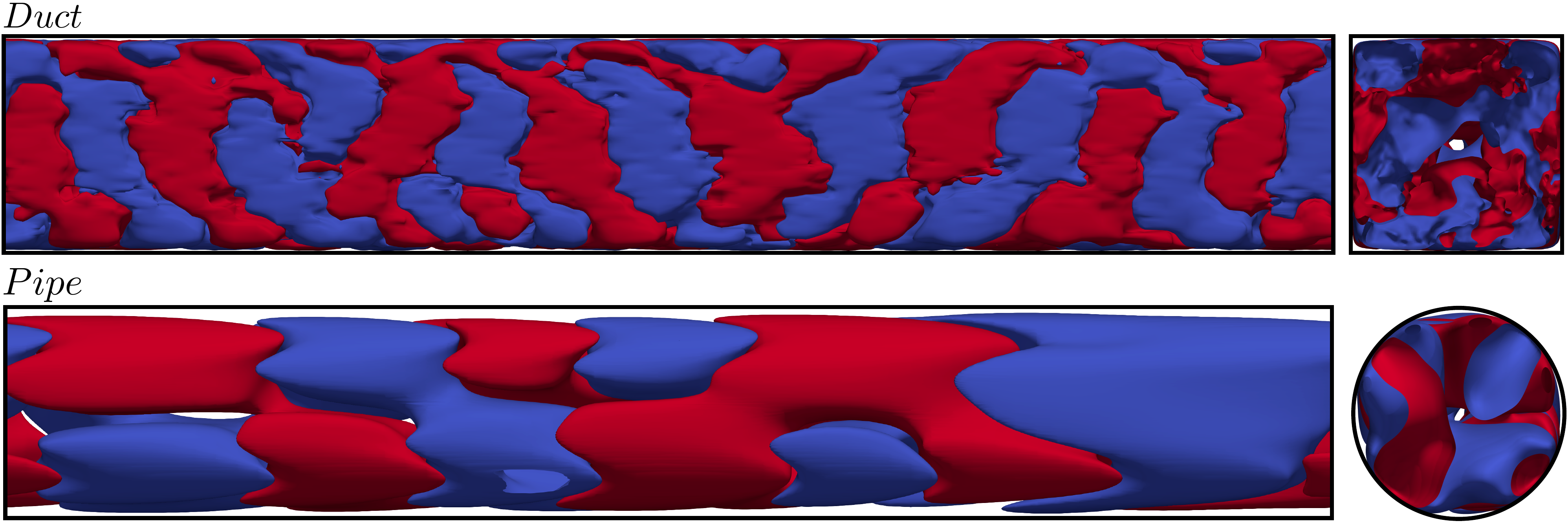}   
   \caption{\textbf{Side and back views of $Q$ iso-surfaces} (left/right) in the duct (top) and pipe (bottom) geometries. $Q$ is defined as the second invariant of the velocity gradient \cite{jeong-hussain-1995}, and its levels are at $Q=\pm 1 \cdot 10^{-4} U^2/\mathcal{L}^2$ in the duct, and $Q=\pm 1.5 \cdot 10^{-3} U^2/\mathcal{L}^2$ in pipe. The flow goes left to right in the side views, and enters the page in the back views.}
   \label{fig:visQ}
\end{figure*}

Denoting by $\mathcal{L}$ either the half-edge of the square duct cross-section or the pipe radius, both geometries extend $4\pi \mathcal{L}$ in the streamwise direction and are discretised on a fine grid with $512$ points streamwise and $1024$ points in each cross-stream direction. When a coarser grid is mentioned, $128$ points streamwise and $256$ points in each cross-stream direction are employed instead. 
{Turbulent statistics reported in the main text are extracted from two orthogonal planes normal to the cross-section,} and do not exhibit significant variations with the grid, as here exemplified for the fluctuating energy spectra (Fig.~\ref{fig:spectraUR}).
The chosen domain size and resolution, in line with previous literature \cite{lellep-linkmann-morozov-2024,foggirota-etal-2024-3}, are sufficient to accommodate and resolve the relevant dynamics of the flow.

Simulations are initialised from fully developed inertial turbulence fields, after which transients are allowed to elapse until a statistically stationary ET state is reached.
All reported measurements are performed in this sustained regime.

\subsection*{Further flow visualisations}

\begin{figure}
   \vspace{-0.5cm}
   \includegraphics[width=\linewidth]{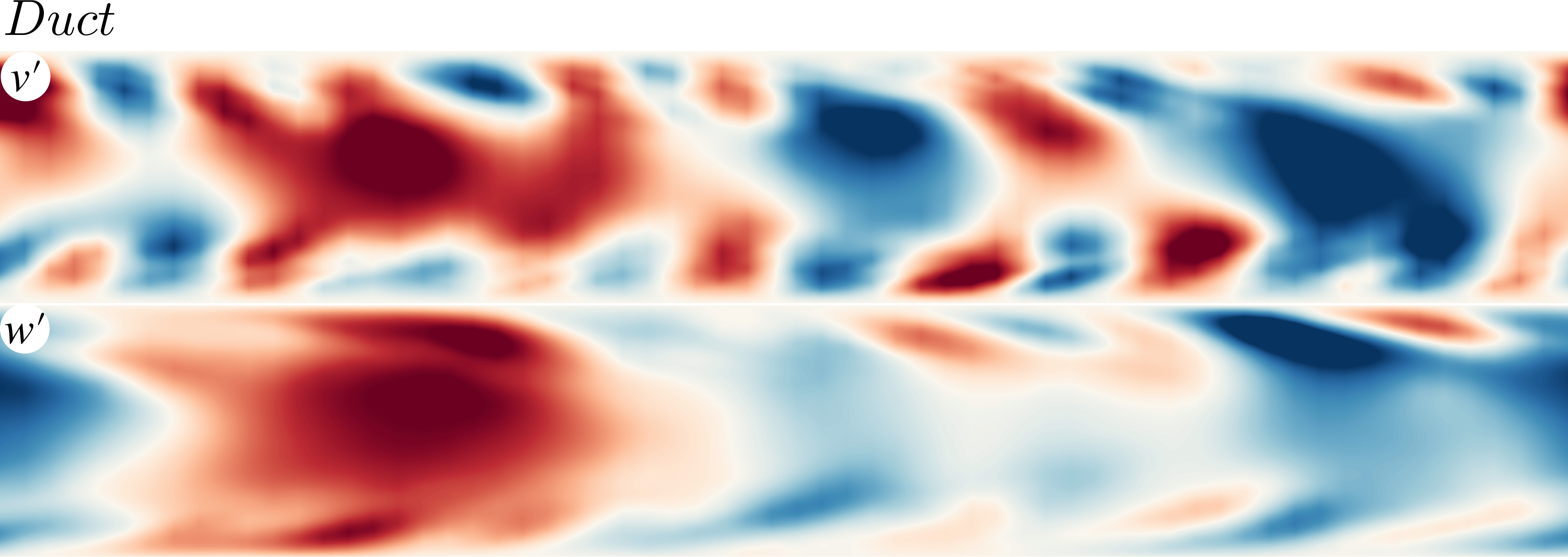}\\[0.3cm]
   \includegraphics[width=\linewidth]{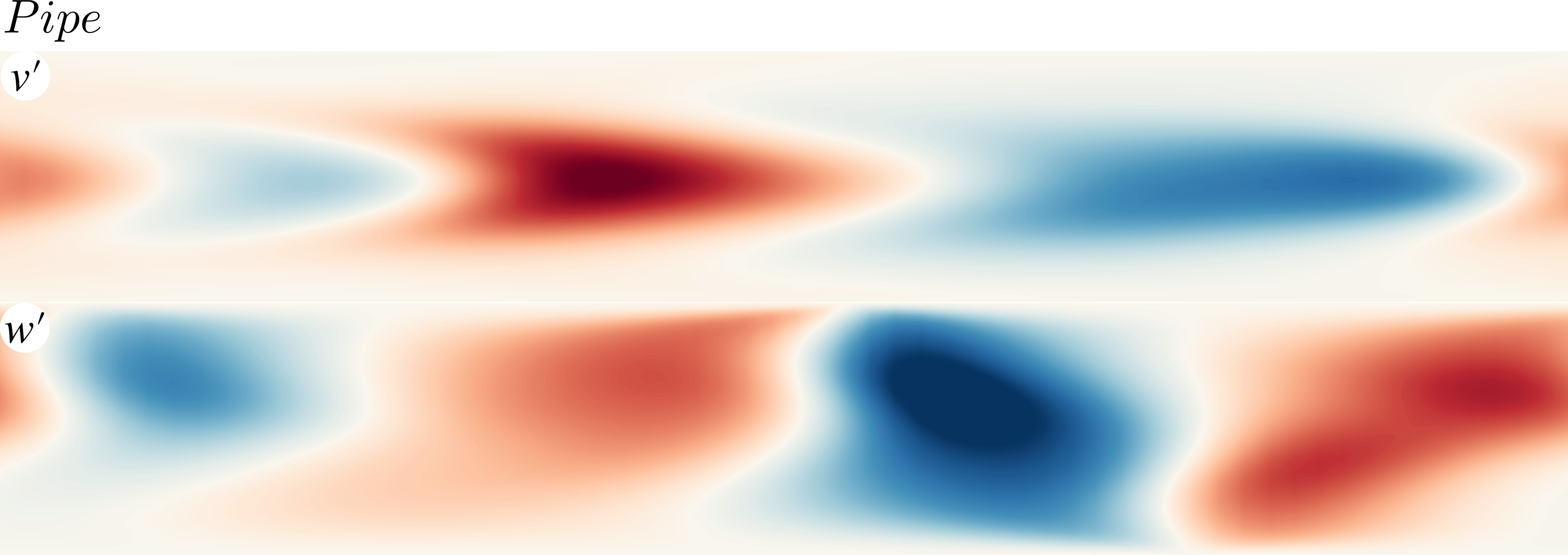}
   \caption{\textbf{Longitudinal slices along the centreline} of the wall-normal ($v^\prime$) and transverse ($w^\prime$) velocity fluctuations.  We adopt a blue to red colormap ranging linearly between $\pm 2 \cdot 10^{-4} U$/$\pm 5 \cdot 10^{-4} U$  for $v^\prime$/$w^\prime$ in the duct and between $\pm 5 \cdot 10^{-2} U$ for both in the pipe. The flow goes from left to right.}
   \label{fig:visSup}
\end{figure}

We consider volumetric visualisations of the $Q$-criterion iso-surfaces (Fig.~\ref{fig:visQ}). These highlight fine-scale activity at the duct walls, and larger-scale motions at the pipe wall, where the breakdown of axial symmetry is also evident. 

Finally, slicing the duct along its centreline (Fig.~\ref{fig:visSup}, top) reveals wall-localised structures reminiscent of the unsaturated PDI \cite{beneitez-page-kerswell-2023,beneitez-etal-2024-2}, together with large-scale coherent fluctuations at the middle suggestive of the centre-mode \cite{garg-etal-2018-1,chaundhary-etal-2021}.
Coherent centreline motions are also clearly observed in the pipe (Fig.~\ref{fig:visSup}, bottom), along with large-scale wall motions evocative of the saturated PDI.

\begin{figure}
   \vspace{-0.3cm}
   \includegraphics[width=\linewidth]{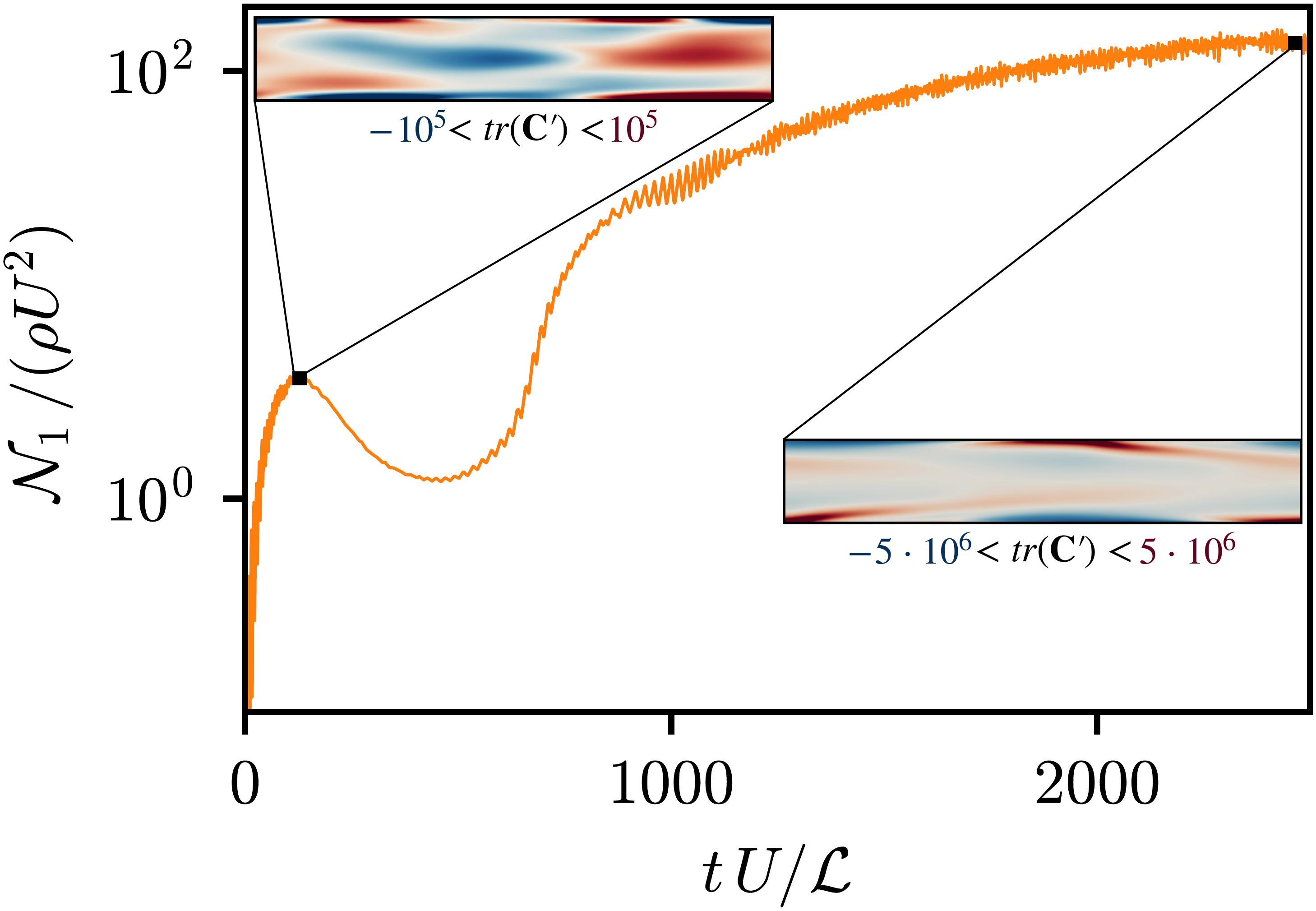}
   \caption{\textbf{Time-history of the first normal-stress difference from our simulation on a coarser grid}, transiently visiting the same state achieved on the fine one before reaching a final state characterised by dominant wall-motions and enhanced stretching of the microstructure.}
   \label{fig:history}
\end{figure}

\subsection*{Increasing diffusivity in the pipe}

Running our pipe simulation on a coarser grid (i.e., increasing the stress diffusivity) leads to a flow state characterised by the same turbulent statistics of the fine one (signature of ET), but a pronounced plug-like velocity profile and modified flow structures. To better characterise this behaviour, we observe the time-history of the first normal-stress difference ($\mathcal{N}_1$) at the pipe centreline throughout the whole extent of the coarser simulation (Fig.~\ref{fig:history}). After the initial transient, the flow momentarily settles into a state characterised by the coexistence of centre- and wall-localised structures: this is the final, sustained state achieved by our simulation on the fine grid.
On the coarser grid, instead, centreline-motions eventually die out while wall-motions intensify, leading to a final state characterised by significantly higher stretching of the microstructure.  

\end{document}